# Anisotropic Infrared Response of Vanadium Dioxide Microcrystals

T. J. Huffman[1], Peng Xu[1], M.M. Qazilbash[1,*], E.J. Walter[1], H. Krakauer[1],
Jiang Wei[2], D.H. Cobden[3], H.A. Bechtel[4], M.C. Martin[4], G.L. Carr[5], D.N.
Basov[6]

1. *Department of Physics, College of William and Mary, Williamsburg, Virginia 23187-8795, USA*
2. *Department of Physics and Engineering Physics, Tulane University, New Orleans, Louisiana 70118, USA*
3. *Department of Physics, University of Washington, Seattle, Washington 98195, USA*
4. *Advanced Light Source, Lawrence Berkeley National Laboratory, Berkeley, California 94720, USA*
5. *Photon Sciences, Brookhaven National Laboratory, Upton, New York 11973, USA*
6. *Department of Physics, University of California, San Diego, La Jolla, California 92093, USA*

Vanadium dioxide ($VO_2$) undergoes a phase transition at a temperature of 340 K between
an insulating monoclinic $M_1$ phase and a conducting rutile phase. Accurate
measurements of possible anisotropy of the electronic properties and phonon features of
$VO_2$ in the insulating monoclinic $M_1$ and metallic rutile phases are a prerequisite for
understanding the phase transition in this correlated system. Recently, it has become
possible to grow single domain untwinned $VO_2$ microcrystals which makes it possible to
investigate the true anisotropy of $VO_2$. We performed polarized transmission infrared
micro-spectroscopy on these untwinned microcrystals in the spectral range between 200
cm$^{-1}$ and 6000 cm$^{-1}$ and have obtained the anisotropic phonon parameters and low
frequency electronic properties in the insulating monoclinic $M_1$ and metallic rutile
phases. We have also performed ab initio GGA+U total energy calculations of phonon
frequencies for both phases. We find our measurements and calculations to be in good
agreement.

## I. INTRODUCTION

Correlated electron systems often possess
multiple interacting degrees of freedom such as
electronic, lattice, magnetic, and orbital. Small
changes in these degrees of freedom can lead to
dramatically different emergent properties such
as high temperature superconductivity, colossal
magnetoresistance, metal-insulator transitions
and multi-ferroicity[1-4]. This leads to
extraordinarily rich phase diagrams as functions
of external parameters such as strain, chemical
doping, applied fields and temperature. Precise
understanding and control of emergent
properties can lead to a number of applications.
However, a complete understanding of how and
why these phase changes occur in complex,
correlated systems has proved elusive.
Experimental measurements provide constraints
based on the observed behavior of complex
systems, thus contributing to our understanding
of these materials.

Vanadium dioxide ($VO_2$) is one such
material[5-8]. $VO_2$ undergoes a metal insulator
transition (MIT) at $T_c$=340K between an
insulating phase below $T_c$ and a conducting
phase above $T_c$. The MIT is accompanied by a
structural transition such that the monoclinic $M_1$
lattice of the insulating phase transforms into the
rutile lattice of the metallic phase. This system
is particularly attractive for the investigation of
strongly interacting degrees of freedom because
of its relatively simple unit cell and because no
chemical doping is required for the phase
transition to occur. Moreover, $VO_2$ shares a
common structural element, a transition metal
inside an oxygen octahedron, with other strongly
correlated systems such as the colossal magneto-
resistive manganites and the high temperature superconducting cuprates. The vanadium ions are in the 4+ valence state, with one electron in the $d$-orbitals[9,10]. Although partially-filled $d$-orbitals would suggest metallic behavior for VO$_2$, it instead exhibits an MIT such that the low temperature ground state is an insulator with an energy gap of ~ 0.6eV[11–15] . The driving mechanism of this MIT is not well understood. There are competing explanations for the MIT in VO$_2$, broadly divided into two categories: the Peierls distortion and the Mott transition. The models based on the Peierls distortion tend to explain the MIT purely in terms of lattice instability, unit-cell doubling, and vanadium-vanadium pairing[9,16]. On the other hand, several researchers, including Mott, have emphasized the role of electronic correlations[17–20].

There are actually three different insulating phases of VO$_2$ (M$_1$, M$_2$, and T). In each of these structures, the displacements of the vanadium atoms from their rutile positions are very different. This indicates that the Peierls distortion may not be sufficient to explain the MIT[17]. Moreover, shifts in optical spectral weight over energy scales of several electron-volts between the monoclinic $M_1$ and rutile phases indicate that correlation effects are indeed important[14]. However, the precise role of the $M_1$ unit cell doubling, whether it is the ultimate cause of the transition or driven by electronic correlations, remains unclear[9,10]. Since the unit cell doubling accompanies the MIT, any explanation of the phase transition in VO$_2$ must account for the structural change. The structural transition may be caused by a softening of an acoustic phonon at the R point[21]. Evidence for this scenario has been provided by x-ray diffuse scattering experiments[22].

The anisotropy of the monoclinic M$_1$ and rutile lattice structures of VO$_2$, along with the inherent anisotropy of the electronic p and d orbitals, may be expected to lead to anisotropy in the electronic and phonon properties. The directional dependence of these properties could play a major role in the MIT. Evidence for the relevance of anisotropy is provided by photoemission and x-ray absorption experiments that have documented the changes in occupation of the a$_{1g}$ and $e_g^\pi$ orbitals across the phase transition[13]. Accurate measurements of the anisotropy of the lattice dynamics and the infrared electronic properties in the monoclinic M$_1$ and the rutile phases are therefore important in the investigation of the cause(s) of the MIT.

Previous infrared spectroscopy experiments on polycrystalline thin films or bulk crystals have been limited in their ability to investigate the anisotropy of VO$_2$ due to the nature of their samples[12,14,23,24]. For example, as large VO$_2$ single crystals go through the structural transition, they exhibit twinning because the symmetry of the rutile (tetragonal) structure is broken in the monoclinic M$_1$ structure. The rutile c$_r$ axis always becomes the monoclinic a$_m$ axis, but only one of the rutile a$_r$ axes can transform to the monoclinic b$_m$ axis leading to two possible orientations of the b$_m$ axis differing by a 90 degree rotation about the c$_r$ axis[24]. Domains approximately 40 micrometers in size result, the difference between the two types of domains being the orientation of the b$_m$ axis[23]. A macroscopic infrared measurement averages over these domains[23]. Therefore, twinning is a problem for measuring the anisotropy of charge dynamics of VO$_2$. Moreover, multi-domain crystals also tend to crack or break as they go through the MIT[23]. As the cracks may introduce new reflection planes, extracting the optical properties from such a crystal is fraught with difficulties. Polycrystalline thin films typically have grains with different orientations. Therefore, in both types of samples, assignment of phonon symmetries and parameters is difficult. Thus either a single domain crystal or an epitaxial, untwinned film is required to make meaningful, specific measurements of the anisotropy of the electronic properties and phonon parameters.

Recently it has become possible to grow untwinned single domain $VO_2$ microcrystals on oxidized silicon substrates by the vapor transport method[25,26]. Single domain crystals are less likely to crack as they go through the MIT. At room temperature, the $a_m$ axis of the monoclinic $M_1$ phase of the microcrystal is parallel to the plane of the substrate[27]. Previously, infrared spectroscopy has been performed on $VO_2$ microcrystals in the spectral range between 1000 $cm^{-1}$ and 7000 $cm^{-1}$. However, these experiments did not consider the anisotropic nature of the $VO_2$ microcrystal and could not measure the infrared active phonons[28]. Broadband infrared microspectroscopy with polarized light allows us to measure a single domain sample to obtain the true anisotropy of the optical constants. In this work, we report the center frequencies, oscillator strengths, and broadenings of 14 of the 15 infrared (IR) active phonons in monoclinic $M_1$ $VO_2$ and all 4 infrared active phonons in rutile $VO_2$, and assign them their proper group theory labels. We compare our results to previous work done on twinned bulk crystals in Ref. 23 and to zone-center frequencies calculated with first principles theory. We also report the directional dependence of the low frequency optical conductivity of metallic rutile $VO_2$ between 200 and 6000 $cm^{-1}$.

The paper is organized as follows: in the next section we present the salient aspects of the experimental methods and data analysis followed by an account of the theoretical methods. Next, we discuss the experimental and theoretical results for the monoclinic phase and then the rutile phase. We conclude by reviewing our main results. Finally, we present technical details about our experiment and data analysis for the experts in the supplemental material[29].

## II. METHODS

### A. Experimental methods

The single domain microcrystals used in this experiment were grown by vapor transport on oxidized silicon substrate[26]. Most of these crystals grow in long, thin rods that are not particularly suitable for infrared microspectroscopy because their narrow dimension tends to be smaller than the diffraction-limit. However, there are a few large microcrystals with low aspect ratios among the ensemble, and we chose one of the largest crystals for our experiment (See Fig. 1a). The thickness of the $VO_2$ microcrystal was directly measured by an atomic force microscope. Layer thicknesses used in the modeling are shown in Fig. 1b. As the substrate is transparent in the infrared spectral range, it is possible to make transmission measurements for obtaining the frequency-dependent complex dielectric function of $VO_2$ microcrystals.

Preliminary characterization of the $VO_2$ microcrystals with infrared microscopy at frequencies greater than 800 $cm^{-1}$ was carried out at the Advanced Light Source (Brookhaven National Laboratory) at Lawrence Berkeley National Laboratory. To extend our spectral range into the phonon region, we later performed broadband infrared micro-spectroscopy between 200 $cm^{-1}$ and 6000$cm^{-1}$ on the $VO_2$ microcrystal at the U12IR beam line at National Synchrotron Light Source. A Fourier Transform Infrared (FTIR) Spectrometer was used to measure the broadband infrared transmission of the $VO_2$ microcrystal and substrate normalized to the transmission of the substrate between 200 $cm^{-1}$ and 6000 $cm^{-1}$. A 15X 0.58 NA Schwarzschild microscope objective focused the FTIR beam onto the microcrystal[30]. A wire-grid on KRS-5 substrate polarizer was used to orient the electric field of the incident light both perpendicular and parallel to the $a_m$ axis of the $M_1$ phase (see Fig. 1c). The orientation of the $a_m$ axis in the $VO_2$ microcrystal was determined by rotating the polarizer until the $A_u$ phonon around 600cm$^{-1}$ was absent in the spectrum[23] . Then, the polarizer was oriented perpendicular to the $a_m$ axis. As the resulting spectrum for $\vec{E} \perp a_m$ contains none of the $B_u$ phonons seen in the $\vec{E}//a_m$ spectrum, it can be concluded that the crystal is oriented such that both the $a_m$ and $b_m$ axes are in the plane of the crystal, i.e. $\vec{E} \perp a_m$ is in fact $\vec{E}//b_m$. In the rutile phase, the incident light was polarized parallel to the $a_r$ and $c_r$ axes (see Fig. 1c). Normalized, broadband transmission spectra were taken at 295K for the monoclinic $M_1$ phase, and at 400K for the rutile phase. The absolute transmission of the substrate was also measured at both these temperatures[29].

Kramers-Kronig consistent oscillators were used to model the normalized transmission spectra and extract $\varepsilon_1$ and $\varepsilon_2$, the real and imaginary parts of the complex dielectric function. Phonon features in the normalized transmission spectra were modeled with Lorentz oscillators of the following form:

$$\varepsilon(\omega) = \sum_{i=1}^{n} \frac{s_i}{1 - \frac{\omega^2}{\omega_i^2} - \frac{i\gamma_i\omega}{\omega_i}}$$

Where $\omega_i$ is the center frequency of the $i^{th}$ phonon in units of inverse wavelength, $s_i$ is the oscillator strength parameter, and $\gamma_i$ is the broadening parameter. The electronic response of the rutile metal was modeled with Lorentz, Tauc-Lorentz and Drude functions.

## B. Theoretical methods

First-principles density functional theory (DFT)[31] calculations were performed using the "Quantum Espresso"[32] computational package, using the DFT+U[33] extension, in order to describe strong V $d$-$d$ orbital correlations. The PBE[34] version of the generalized gradient approximation (GGA) was used for all calculations and only non-magnetic ground states were considered throughout this work. The rotationally invariant[35] form of the GGA+U approach is used to apply the Hubbard U correction. For all systems, we investigated a range of U corrections ranging from 0-7 eV. A Hubbard U value of 5 eV was found to give good agreement for both structures, as discussed further below. Lattice parameters for rutile[36] and monoclinic $M_1$[37] structures were fixed at values obtained from x-ray diffraction measurements. All internal atomic coordinates were relaxed until the calculated forces were less than 1 mRy/Bohr (~ 0.03 eV/Angstrom). Ultrasoft pseudopotentials[38] were obtained from the Quantum Espresso website for vanadium and oxygen[39]. Tests showed that a wave function planewave cutoff of $E_{cut}$ = 50 Ry and a charge density cutoff of 300 Ry was sufficient to yield converged total energies and forces. Brillouin zone integrations were performed using 6x6x8 and a 4x4x4 Monkhorst-Pack[40] k-point meshes for the rutile and monoclinic structures, respectively; a small Fermi-Dirac type temperature broadening of 0.05 eV was also used in the metallic rutile phase. Zone center phonons were calculated using the method of small displacements and analyzed using the "Phonopy"[41] program, and Born effective charge tensors Z* and $\varepsilon_\infty$ were used to include non-analytic contributions to the dynamical matrix.

## III. RESULTS AND DISCUSSION

### A. Monoclinic $M_1$ phase

There are 12 atoms in the unit cell of monoclinic $M_1$ $VO_2$ of space group P2$_1$/c[9]. Group theory then demands that there will be 36 phonon modes, of which 3 are acoustic, 18 are Raman active, and 15 are IR active. The longitudinal or transverse character of a particular IR active mode depends upon the direction of the phonon wave-vector $q$. When $q$ is along the $b_m$ axis, the 8 $A_u$ IR modes have purely longitudinal character, while the 7 $B_u$ IR modes have purely transverse character[42]. That is to say that the net dipole moments are along $b_m$ for the $A_u$ modes and perpendicular to $b_m$ for the $B_u$ modes. Along other wave vectors the phonons will have mixed transverse/longitudinal character due to the low symmetry of the $M_1$ phase. In the present experiment, the wave vector is perpendicular to the $a_m$-$b_m$ plane so all 15 IR modes are expected to be seen: 8 $A_u$ modes when the light is polarized along $b_m$ and the 7 $B_u$ modes when the light is polarized along $a_m$ (perpendicular to $b_m$).

Experimentally, we see 7 distinct phonon features when the electric field ($\vec{E}$) of the incident light is parallel to $a_m$ ($B_u$), and 7 distinct features when $\vec{E}$ is parallel to $b_m$ ($A_u$) (See Fig. 2). The eighth $A_u$ feature, which has been seen by Barker et al.[23] at 189 cm$^{-1}$ in a bulk, twinned crystal, is outside of our spectral range. All 8 $A_u$ modes are thus accounted for. Table 1 tabulates the measured phonon parameters. The $A_u$ peak near ~ 600 cm$^{-1}$ in Fig. 2 is asymmetric, which required two Lorentz oscillators, $\omega$ = 607 and 637 cm$^{-1}$, to fit, as shown in Table 1. We speculate that the apparent asymmetry observed near ~ 600 cm$^{-1}$ could be due to two-phonon processes arising from phonons near ~ 300 cm$^{-1}$.

In general, our $A_u$ center frequencies are in good agreement with Ref. 23 (See Fig. 3). Whereas Ref. 23 only identifies one mode at 505 cm$^{-1}$, our increased spectral resolution of 2 cm$^{-1}$, as opposed to the ~7.5 cm$^{-1}$ resolution of Ref. 23, allows us to resolve two distinct features at 500 cm$^{-1}$ and 521 cm$^{-1}$, so that we see 7 IR active $A_u$ modes within our measured spectral range. The broadenings of the $A_u$ modes are within 10% of those measured by Ref. 23 for comparable modes. It should be noted that the $A_u$ parameters in Ref. 23, measured as $E \perp a_m$, were extracted from a spectrum with all 15 IR active phonons due to the twinning of the bulk crystal. Our $B_u$ ($\vec{E}//a_m$) mode center frequencies largely agree with Ref. 23. However, we resolve two distinct modes at 351 cm$^{-1}$ and 367 cm$^{-1}$ whereas Ref. 23 reports a single mode at 355 cm$^{-1}$. With the inclusion of this mode, all 7 $B_u$ modes are accounted for in our data. Ref. 23 also reports very weak modes at 227.5 cm$^{-1}$ and 478 cm$^{-1}$, which do not appear in our data. For comparable modes, the broadenings for the $B_u$ modes agree reasonably well with those of Ref. 23, with the exception of $B_u$ mode 6, which is nearly twice as broad in the present work. It should be noted that this discrepancy is due in part to the "extra" mode used in Ref. 23 at 478 cm$^{-1}$. Moreover, the larger broadening of $B_u$ mode 6 could be due to contribution from two phonon processes associated with $B_u$ mode 1. Even though the $a_m$ axis is in the same direction for all the domains in Ref. 23's twinned crystal, the phonon parameters, particularly the oscillator strengths, will depend on the orientation of the wave vector of the incident light relative to the $b_m$ and $c_m$ axes. Thus, the oscillator strengths between our measurements on an untwinned crystal and those of Ref. 23 cannot be directly compared.

The phonon center frequencies at the gamma point have been calculated using GGA+U calculations. A Hubbard U = 5 eV yields good agreement with our experimental frequencies for both structures as seen in Table 3. Furthermore, this value yields a band gap of ~ 1.0 eV which is in line with the experimentally determined band gap of ~0.6 eV[11–15]. In general, there will be a frequency shift in the IR modes due to LO/TO-type splitting. These require knowledge of the high-frequency dielectric constant $\varepsilon_\infty$ as well as the Born effective charge tensors Z*. Due to technical complications in calculating $\varepsilon_\infty$ and $Z^*$ with GGA+U, we used the experimentally determined value $\varepsilon_\infty \sim 12$ from the present work; two sets of $Z^*$ tensors were used, the first from monoclinic $ZrO_2$ from Ref. 42 and, for comparison, the second using nominal diagonal values $Z_{\alpha\beta}^*(V)=+5e\delta_{\alpha\beta}$ and $Z_{\alpha\beta}^*(O)=-2.5e\delta_{\alpha\beta}$ [$Z_{\alpha\beta}^*(O)$ were simply fixed using the acoustic sum rule]. For the sample geometry, photon wave vectors are perpendicular to the $a_m$-$b_m$ plane (the $\Gamma$ to Y direction in the Brillouin zone), so only the $B_u$ frequencies depend on $\varepsilon_\infty$ and $Z^*$[42].

As seen in Table 3, the agreement between experiment and theory is slightly better for the $B_u$ modes than for the $A_u$ modes. The computed mean absolute deviation (MAD) for the $A_u$ modes is 45 cm$^{-1}$ while for the $B_u$ modes it is 39 cm$^{-1}$ or 35 cm$^{-1}$ depending on the choice of Born tensors: values in parenthesis are for the nominal $Z^*$ discussed above. The agreement between our measured and computed frequencies (and the band gap) can be improved by decreasing the Hubbard U correction to 3 - 4 eV. However, this greatly increases the discrepancy for rutile (see next section).

## B. Rutile phase

The rutile structure is a simple tetragonal unit cell containing 6 atoms with space group P4$_2$/mnm[36,43] . Group theory predicts that there will be 18 rutile $VO_2$ phonons: 3 acoustic, 3 silent, 5 Raman active and 7 infrared active modes. Of the 7 infrared active phonons, 3 are doubly degenerate. Thus, we expect to see 1 $A_{2u}$ mode when $\vec{E}//c_r$, and 3 $E_u$ modes when $\vec{E}//a_r$. Experimentally, all four rutile infrared active phonons of $VO_2$ are seen for the first time. The measured phonon parameters and the low frequency optical conductivity are shown in Table 2 and Fig. 4 respectively. The rutile $VO_2$ phonons are roughly three times as broad as those of the monoclinic $M_1$ phase and insulating rutile $TiO_2$[44]. This broadening implies a decrease in phonon lifetime possibly due to electron-phonon coupling.

Phase coexistence in the form of a stripe pattern with alternating insulating and metallic regions in the microcrystals was observed through an optical microscope in the MIT regime. Similar stripe patterns have previously been observed in $VO_2$ nano-rods. These stripes have been shown to be phase domains that form due to stress in the microcrystals caused by mismatch of thermal expansion between the oxidized silicon substrate and the $VO_2$ microcrystal[26,45]. As the microcrystals are grown at 1273 K, the mismatch between the coefficients of thermal expansion results in a ~ 0.7% in-plane isotropic tensile strain on the rutile structure near the phase transition temperature. The $M_1$ structure then expands by ~1.1% along the $a_m$ axis during the phase transition from rutile to monoclinic, whereas there is little change along the $b_m$ axis. Thus, the monoclinic phase is under a ~ 0.4% compressive strain along the $a_m$ axis, and ~ 0.7% tensile strain along the $b_m$ axis[26,45,46]. This leads to a monoclinic $M_1$ unit cell volume that is only slightly larger than that of bulk $VO_2$. In this way, it is possible that the effects of strain on the monoclinic $M_1$ phase are minimized, as our center frequencies are in good agreement with those obtained by Ref. 23 on bulk $VO_2$. Strain effects could play a larger role in the rutile phase properties.

The overall shape of the electronic conductivity between 2000 cm$^{-1}$ and 6000 cm$^{-1}$ for the rutile metal (Fig. 4) is consistent with previous reports[12,14,24]. The overall higher conductivity along the rutile $c_r$ axis compared to the $a_r$ axis is consistent with Ref. 24. Optical conductivity below 2000 cm$^{-1}$ along the $a_r$ and $c_r$ axes of the rutile phase has not been previously reported in the literature. Our data, which extends down to 200 cm$^{-1}$, suggests that the higher conductivity along the $c_r$ axis compared to the $a_r$ axis persists down to these frequencies. This is consistent with *dc* resistivity measurements made on single crystal VO$_2$[47]. However, the degree of anisotropy was much greater than that seen in our experiment, as the dc conductivity parallel and perpendicular to $c_r$ was reported to be 2500 $\Omega^{-1}$cm$^{-1}$ and 333 $\Omega^{-1}$cm$^{-1}$ respectively[47]. An even greater degree of anisotropy is seen in highly strained VO$_2$ thin films on TiO$_2$ substrates, where the dc conductivity is measured to be 41.5 times greater along $c_r$ than along $a_r$. These films are under a 1.92% tensile strain along $c_r$ and a .93% compressive strain along $a_r$[48]. Thus, the anisotropy of the conductivity at low frequencies is very sensitive to strain. Below 1000 cm$^{-1}$, there is uncertainty in our conductivity data due to the incident spot being larger than the sample. The uncertainty, discussed in greater detail in the supplemental material[29], is large enough to preclude definitive statements about the frequency dependence of the electronic response at low frequencies. The possibility of localization of the conduction electrons, as evidenced by a peak in the conductivity ($\sigma_1$), around 1500 cm$^{-1}$ in both polarizations is within the experimental uncertainty. A similar peak in $\sigma_1$ has been seen previously in nano-scale metallic "puddles" of VO$_2$ near the phase transition[49,50]. However, such a feature has not been seen in previous macroscopic experiments on polycrystalline thin films of VO$_2$[12,14].

The agreement between experiment and theory for the phonon frequencies of the rutile structure is on par with the M$_1$ results (See Table 4). The MAD for the rutile modes is 41 cm$^{-1}$. Interestingly, the 3 E$_u$ modes are still in good agreement for U = 3 eV (their errors are 17, 59, and 61 cm$^{-1}$), however, the A$_{2u}$ mode is unstable (large negative $\omega^2$) until U is increased to about

5 eV. In rutile TiO$_2$, this mode is associated with an incipient ferroelectric phase; under negative pressure, calculations show that it softens, resulting in a ferroelectric phase transition[51]. For values of U smaller than about 5 eV in rutile VO$_2$, the same ferroelectric-like instability incorrectly appears at the experimental volume. Note that in VO$_2$ it is not a true ferroelectric state, since the system remains metallic when the crystal is allowed to distort according to this mode.

As mentioned, microcrystals in the rutile phase are under ~ 0.7 % tensile strain along the $a_r$ axis. To examine strain effects, we recomputed the phonon frequencies in the presence of -1% strain along the $c_r$ axis, relaxing the in-plane axes. Differences between calculated and measured phonon frequencies changed by less than ~ 9 cm$^{-1}$, except for the second highest E$_u$ mode which increased by 22 cm$^{-1}$. Strain effects of this magnitude are thus not likely to be responsible for the differences between theory and experiment.

## IV. CONCLUSIONS

Polarized infrared micro-spectroscopy of untwinned single domain VO$_2$ crystals was performed. Single domain samples allow for the measurement of the true anisotropy of the phonons and the electronic response. The four zone-center infrared active phonons of metallic rutile VO$_2$ have been measured and identified for the first time[29]. The electronic part of the infrared conductivity of metallic rutile VO$_2$ is weakly anisotropic and is measured to be higher along the $c_r$ axis as compared to the $a_r$ axis. The oscillator parameters of 14 of the 15 zone center infrared active phonon modes of the monoclinic M$_1$ phase of untwinned VO$_2$ have been measured for the first time[29]. In addition, we have resolved an A$_u$ mode near 500 cm$^{-1}$ and observe a distinct B$_u$ mode at 367 cm$^{-1}$ not seen in previous measurements reported in Ref. 23.

From our measurements together with the lowest frequency $A_u$ mode seen in Ref. 23, all 15 monoclinic $M_1$ infrared active phonons are now accounted for. Using first-principles GGA+U calculations we have computed the zone-center phonon frequencies for monoclinic and rutile $VO_2$. Our calculated results agree well with our measured frequencies.

## V.    SUPPLEMENTAL MATERIAL

It is necessary to account for the fact that, at long wavelengths, the microscope spot size is larger than the crystal. It is natural to model this situation in the following way.

$$T_{Eff} = F(\omega)T_{VO_2+Sub} + (1 - F(\omega))T_{Sub}$$

Where $F(\omega)$ is the percentage of light first incident on the $VO_2$ at a given frequency, $T_{VO_2+Sub}$ is the absolute transmission through the $VO_2$ and substrate, and $T_{Sub}$ is the absolute transmission through the substrate. $\frac{T_{Eff}}{T_{Sub}}$ is the data that was actually measured. $\frac{T_{Eff}}{T_{Sub}}$ is equivalent to $\frac{T_{VO_2+Sub}}{T_{Sub}}$ at higher frequencies, when all of the light is falling on the crystal At low frequencies it is necessary to use $F(\omega)$ to extract $\frac{T_{VO_2+Sub}}{T_{Sub}}$ from $\frac{T_{Eff}}{T_{Sub}}$, as $\frac{T_{VO_2+Sub}}{T_{Sub}}$ is the real data of interest.

The size and shape of the sample and the spot are the same between the two phases. Thus, $F(\omega)$ is the same for all phases and polarizations. This assumes that the transmission is a simple geometric sum of the infrared light passing through the crystal and that passing through the area around it. This ignores potential scattering and plasmonic effects due to the material's polarizability and shape. The lack of any strong dispersive features in the transmission spectra supports the notion that scattering and plasmonic effects are not significant.

Because of the complex shape of the measured crystal, and the presence of surrounding crystals of different thicknesses, $F(\omega)$ cannot be calculated precisely. The error bars in Fig. 4 are due to this uncertainty. The known rutile $c_r$ $dc$ conductivity, the known spot profile of the objective, and the size of the measured microcrystal, were used to constrain $F(\omega)$. For a 0.58NA Schwarzschild objective, the spot size starts to exceed 50 microns at 1000 $cm^{-1}$ (See Fig. S1) and becomes larger than the size of the crystal. Thus, we used an $F(\omega)$ equal to 1 above 1000 $cm^{-1}$. In the low frequency limit, the transmission through a film depends only on the $dc$ conductivity[52]. The $dc$ conductivity along $c_r$ for nano-rods is known to be 2000 $\Omega^{-1}cm^{-1}$ [26]. To make the measured data consistent with this constraint, $F(\omega)$ must be 0.77 at 200 $cm^{-1}$. The intermediate values of $F(\omega)$ were then interpolated with a parabolic curve connecting the two constraining points at 200 $cm^{-1}$ and 1000 $cm^{-1}$, with the vertex of the parabola at ($\omega$ =1000 $cm^{-1}$, $F(\omega)$=1).

If the $dc$ conductivity constraint is lifted, $F(\omega)$ is still constrained in that it must return a positive transmission in rutile $VO_2$. This sets a lower limit on $F(\omega)$ at 200 $cm^{-1}$; $F(200\ cm^{-1})$ must be greater than 0.68. However, this would yield a very large $\sigma_1(\omega)$ in the $dc$ limit. A more reasonable lower limit for $F(200\ cm^{-1})$ of 0.72 was considered as it yields a $dc$ conductivity along $c_r$ of 3000 $\Omega^{-1}cm^{-1}$, 50% larger than that measured by Ref. 26.

Finite element analysis was used to check the reliability of the $F(\omega)$ values. The optical image of the sample was discretized, and $F(\omega)$ was calculated numerically using the following formula.

$$F(\omega) = \frac{\int_{microcrystal} I(\omega, x, y)dxdy}{\int_{All\ Space} I(\omega, x, y)dxdy}$$

Where $I(\omega\, x, y)$ is the intensity of the spot at a given position for a specific frequency. While the different thicknesses of the surrounding crystals make an exact calculation of $F(\omega)$ impossible, it is useful in that it can establish an upper limit on $F(200\ cm^{-1})$. We note that the surrounding crystals are much thicker than the crystal being measured and therefore contribute an insignificant amount to the measured transmission spectra.

A spatial aperture was used during the experiment to "apodize" the canonical Schwarzchild intensity distributions shown in Fig. 5, reducing the intensity in the higher order rings at the cost of increased width of the central maximum[30], thereby ensuring that the crystal being measured contributes overwhelmingly to the transmission spectra at the expense of the surrounding crystals. Thus, in actuality, $F(200\ cm^{-1})$ will be greater than what would be implied by the intensity distributions in Fig. 5. Supposing that all of the light falls within the first order minima, finite element analysis yields an $F(200\ cm^{-1})$ of 0.97. Likewise, assuming that all of the light falls within the second order minima yields an $F(200\ cm^{-1})$ of 0.63. While 0.97 is unreasonably high, and 0.63 is less than the physical limit of 0.68 discussed above, averaging these values and adding an extra 5% provides a reasonable upper limit on $F(200\ cm^{-1})$ of 0.85. Figure 6 shows the uncertainty in the optical constants due to the uncertainty in $F(\omega)$ using the above considerations to bound $F(200\ cm^{-1})$ between 0.72 and 0.85.

The normalized transmission spectra were modeled using WVASE 32, spectroscopic analysis software from J.A. Woollam Co. The microscope objective leads to a range of angles of incidence from 15 degrees to 35 degrees. The angle of incidence was modeled as 25 degrees. As the variance in the modeled transmission between angles of incidence between 15 and 35 degrees is less than 1%, this approximation has a negligibly small effect on the modeling. Spectra and fits are shown in Fig. 7.

The rutile transmission spectra in the phonon region warrant special comment. The phonon features in the rutile phase spectra are much weaker than those of the monoclinic $M_1$ phase. However, there are four reproducible features above our noise level, one when $\vec{E}//c_r$, and three when $\vec{E}//a_r$. Fig. 7 (e) and (f) show the data, and the generated transmission with the phonon oscillators (grey) and without the phonon oscillators (red) present in the model. It is clear that while the features are weak, the phonon oscillators are necessary to achieve an acceptable fit.

## VI. ACKNOWLEDGEMENTS

The National Synchrotron Light Source is supported by the U.S. Department of Energy under contract DE-AC02-98CH10886. D. N.B. acknowledges support from the US Department of Energy. E.J.W. acknowledges support by ONR grants N000140910300 and N000141110563. H.K. acknowledges support by ONR grants N000140811235 and N000141211042. The work in M.M.Q.'s group was partly supported by the Jeffress Memorial Trust and by a grant from NSF DMR (1255156).

*Correspondence should be addressed to M.M.Q. at mumtaz@wm.edu

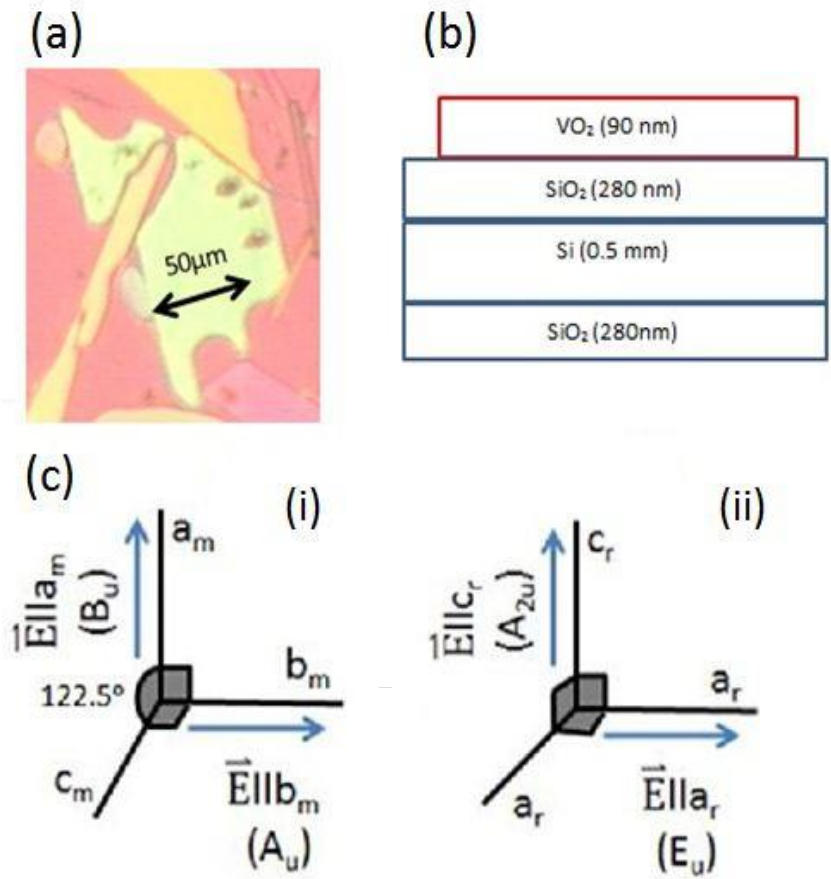

FIG. 1. (a) Optical image of the VO$_2$ microcrystal studied. The double-ended arrow indicates the size of the crystal. (b) Schematic cross sectional view of the microcrystal and substrate. The substrate is oxidized silicon. The thicknesses of the various layers are given in parentheses. (c) Diagram of the polarizations used with respect to the crystallographic axes in (i) the monoclinic (M$_1$) phase and (ii) the rutile phase.

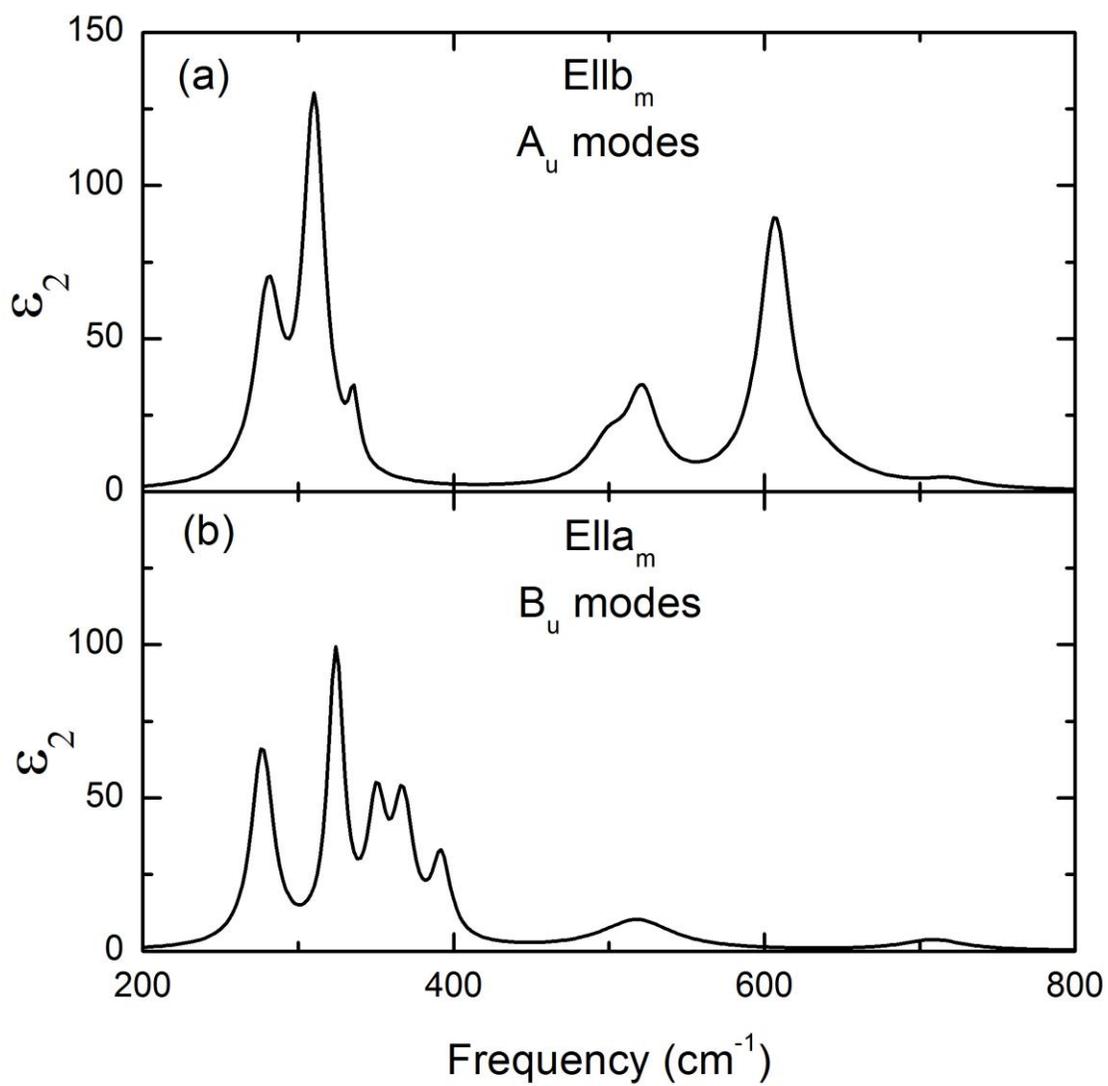

FIG. 2. The experimentally derived imaginary part of the complex dielectric function for monoclinic ($M_1$) VO$_2$ at T=295K for (a) $\vec{E}//b_m$ (A$_u$) and (b) $\vec{E}//a_m$ (B$_u$) in the phonon spectral region.

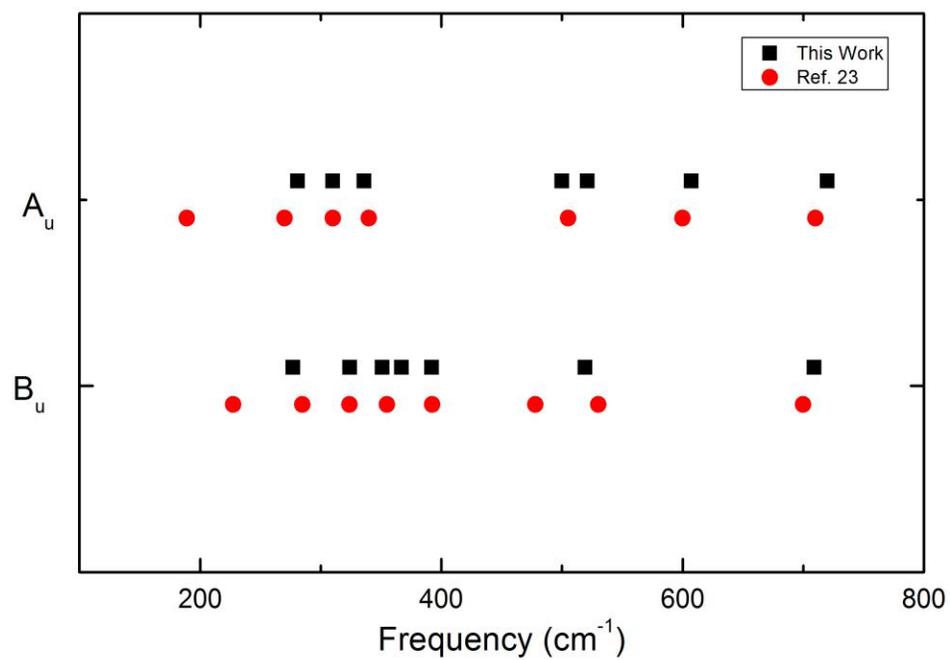

**FIG. 3. Comparison of VO$_2$ monoclinic ($M_1$) center frequencies of A$_u$ and B$_u$ phonon modes from our experiment and Ref. 23.**

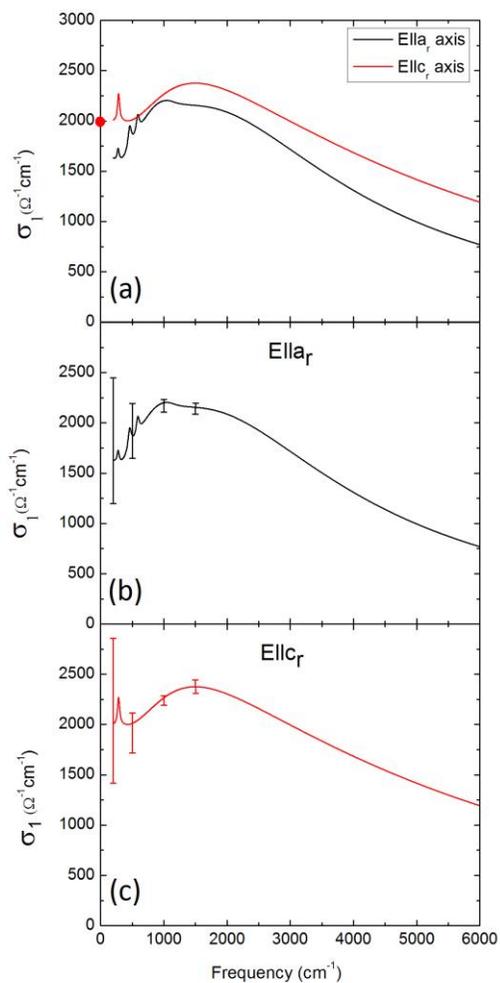

FIG. 4. (a) The $a_r$ axis and $c_r$ axis infrared conductivity ($\sigma_1$) of rutile $VO_2$ at T=400K. The plots are consistent with the $c_r$ axis *dc* conductivity constraint explained in the supplemental material[29]. The known dc conductivity along $c_r$ is shown by the red circle in (a)[26]. Lifting this constraint produces the error bars shown in Fig. 4 (b) and (c). These error bars arise from systematic uncertainties explained in the supplemental material[29]; the systematic uncertainties affect the conductivity ($\sigma_1$) of both axes in a similar manner.

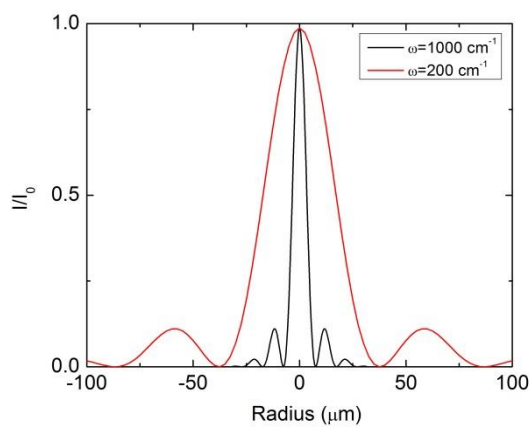

**FIG. 5. Intensity profile of a 0.58 NA Schwarzschild objective at two representative frequencies.**

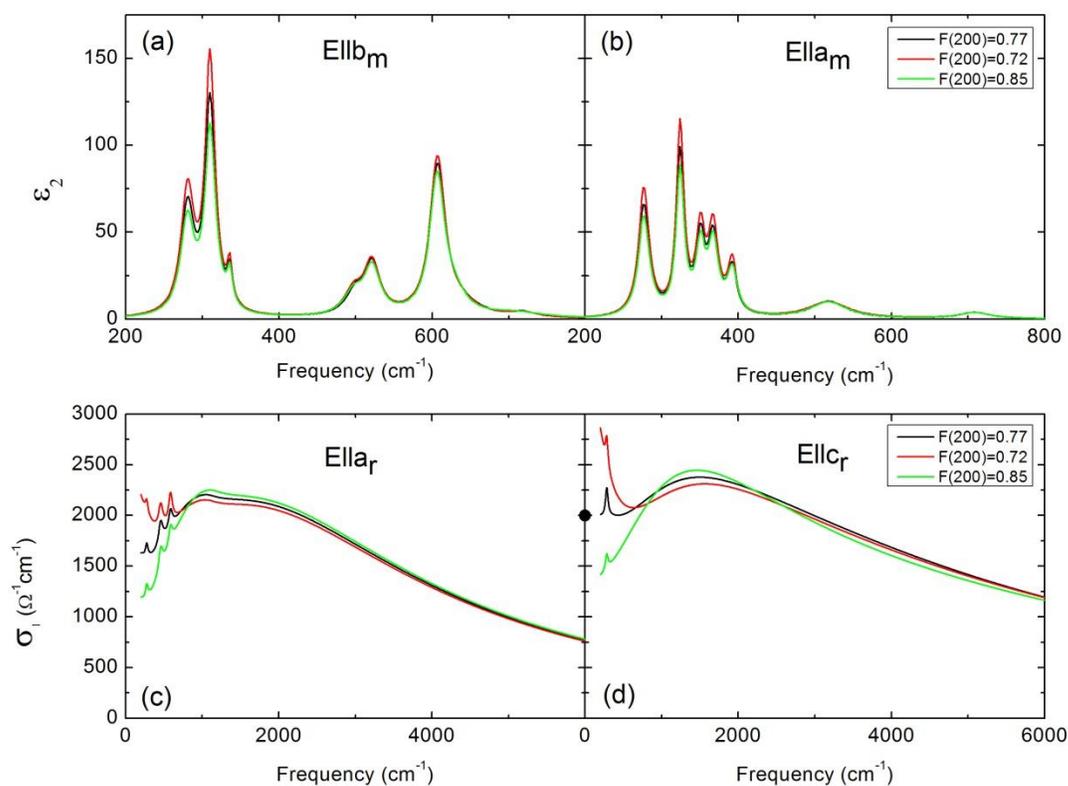

**FIG. 6. The uncertainty caused by the spread in F(ω) for monoclinic *(M₁)* VO₂ (a) and (b), and rutile VO₂ (c) and (d). Black curves show the optical constants consistent with the rutile cᵣ axis *dc* conductivity constraint (black circle). Red curves show the optical constants using the lower limit on *F(200 cm⁻¹)*. Green curves show the optical constants using the upper limit *on F(200 cm⁻¹)*.**

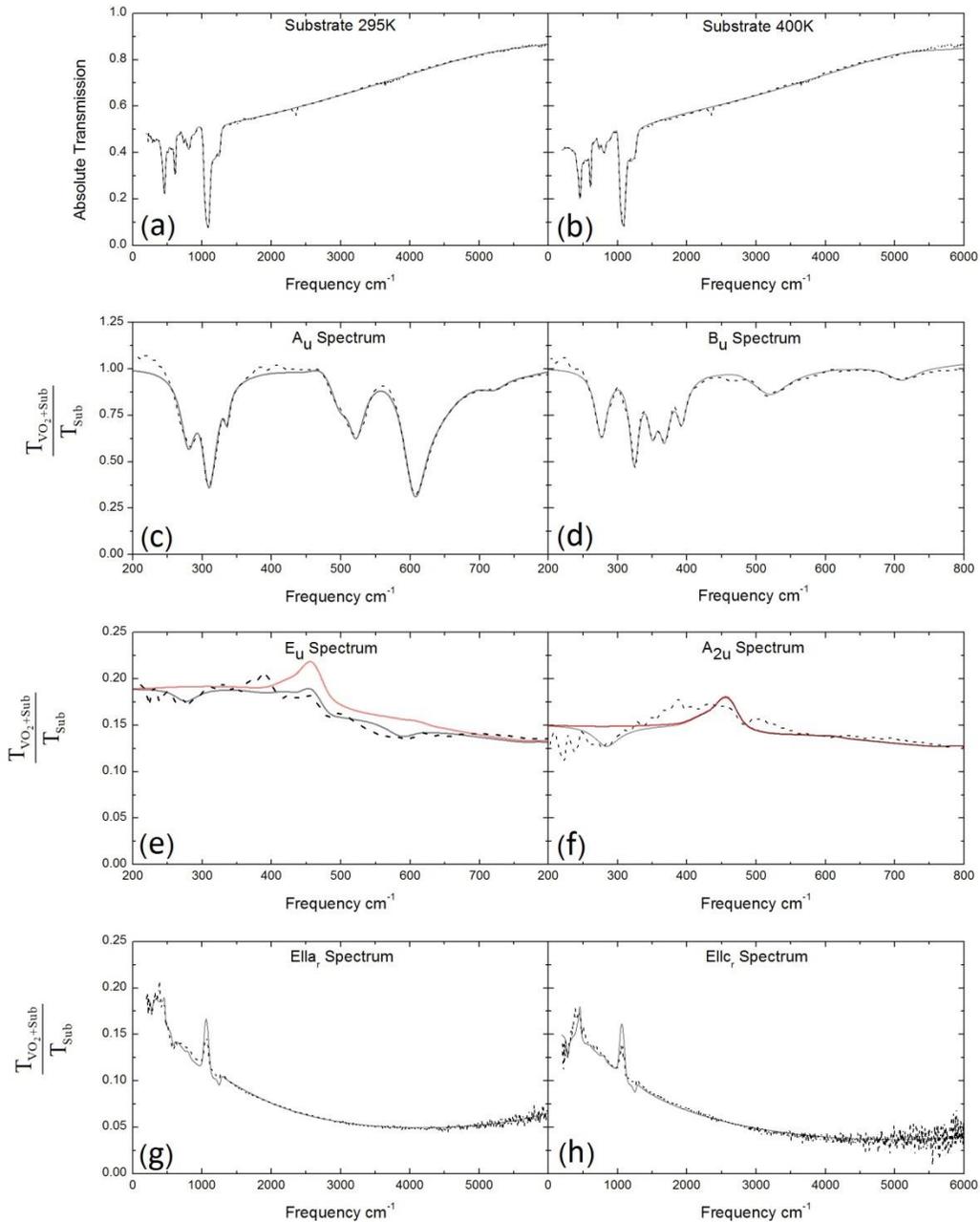

FIG. 7. Representative spectra and fits for the substrate at (a) 295K and (b) 400K; monoclinic ($M_1$) VO$_2$ at 295K (c) and (d); and rutile VO$_2$ at 400K (e), (f), (g) and (h). Data are shown as dashed lines and fits are shown as solid lines. The red lines in (e) and (f) show the modeled transmission without the phonon oscillators. Data shown in (c), (d), (e), (f), (g) and (h) is consistent with $F(\omega)$ obtained from the rutile $c$, axis dc conductivity constraint (F(200)=0.77).

**Table 1. Lorentz oscillator fit parameters for monoclinic (M$_1$) VO$_2$ zone-center infrared active phonons**

| Mode no. | A$_u$ modes | | | B$_u$ modes | | |
|---|---|---|---|---|---|---|
| | $\omega_i$ (cm$^{-1}$) | $s_i$ | $\gamma_i$ | $\omega_i$ | $s_i$ | $\gamma_i$ |
| (1) | (189) | (0.54) | (0.012) | 277 | 4.01 | 0.062 |
| 2 | 281 | 4.53 | 0.074 | 324 | 3.49 | 0.038 |
| 3 | 310 | 6.69 | 0.055 | 351 | 1.67 | 0.041 |
| 4 | 336 | 0.49 | 0.023 | 367 | 1.88 | 0.044 |
| 5 | 500 | 0.77 | 0.060 | 392 | 0.99 | 0.038 |
| 6 | 521 | 1.34 | 0.047 | 519 | 1.08 | 0.110 |
| 7 | 607 | 3.42 | 0.040 | 709 | 0.25 | 0.071 |
| | 637 | 0.67 | 0.100 | - | - | - |
| 8 | 720 | 0.15 | 0.056 | - | - | - |

**Notes: A$_u$ mode 1 is from Ref. 23 as it falls just outside our spectral range. A$_u$ mode 7 is asymmetric and requires two oscillators for a proper fit.**

**Table 2. Lorentz oscillator fit parameters for rutile VO$_2$ zone- center infrared active phonons**

| Mode no. | A$_{2u}$ mode | | | E$_u$ modes | | |
|---|---|---|---|---|---|---|
| | $\omega_i$ (cm$^{-1}$) | $s_i$ | $\gamma_i$ | $\omega_i$ | $s_i$ | $\gamma_i$ |
| 1 | 284 | 8.33 | 0.141 | 277 | 4.12 | 0.148 |
| 2 | - | - | - | 460 | 4.65 | 0.152 |
| 3 | - | - | - | 588 | 1.88 | 0.103 |

Table 3. Comparison of experimental and theoretical phonon frequencies for monoclinic (M₁) VO₂. The Mean Absolute Difference (MAD) between the theory and experiment is given for both phonon symmetries. For the B$_u$ theory values the non-analytic correction includes $ZrO_2$ Born effective charges (see text), whereas the frequencies in parentheses used nominal charges for V and O.

| Mode no. | A$_u$ mode center frequencies (cm$^{-1}$) | | | B$_u$ mode center frequencies (cm$^{-1}$) | | |
|---|---|---|---|---|---|---|
| | Experiment | Theory | Difference | Experiment | Theory | Difference |
| 1 | (189) | 149 | 40 | 277 | 218 (227) | 59 (50) |
| 2 | 281 | 246 | 35 | 324 | 292 (292) | 32 (32) |
| 3 | 310 | 275 | 35 | 351 | 370 (370) | -19 (-19) |
| 4 | 336 | 355 | -19 | 367 | 403 (402) | -36 (-35) |
| 5 | 500 | 417 | 83 | 392 | 466 (434) | -74 (-42) |
| 6 | 521 | 466 | 55 | 519 | 544 (551) | -25 (-32) |
| 7 | 607 | 512 | 95 | 709 | 738 (754) | -29 (-45) |
| 8 | 720 | 720 | 0 | - | - | - |
| | | | MAD 45cm$^{-1}$ | | | MAD 39 (36) cm$^{-1}$ |

Notes: A$_u$ mode 1 is from Ref. 23 To compare with theory, 607cm$^{-1}$ is used as the center frequency for A$_u$ mode 7, as it is the center frequency of the stronger of the two oscillators used to model A$_u$ mode 7 (See Table 1). B$_u$ theoretical values are for Z$^*$ taken from $ZrO_2$ in Ref. 42, while values in parenthesis are for nominal Z$^*$ values (see text).

Table 4. Comparison of experimental and theoretical phonon frequencies for rutile VO₂. The mean absolute difference (MAD) between theory and experiment is given for all rutile phonons.

| Mode no. | A$_{2u}$ mode center frequency (cm$^{-1}$) | | | E$_u$ mode center frequencies (cm$^{-1}$) | | |
|---|---|---|---|---|---|---|
| | Experiment | Theory | Difference | Experiment | Theory | Difference |
| 1 | 284 | 269 | 15 | 277 | 215 | 62 |
| 2 | - | - | - | 460 | 398 | 62 |
| 3 | - | - | - | 588 | 563 | 25 |
| | | | MAD 41cm$^{-1}$ | | | |